\documentclass[12pt]{article}

\def\a{\alpha}

\def\b{\beta}

\def\Th{\Theta}

\def\s{\sigma}
\def\f{\phi}

\def\r{\rho}

\def\G{\Gamma}
\def\g{\gamma}

\def\d{\delta}

\def\L{\Lambda}

\def\t{\tau}

\def\cv{{\cal V}}

\def\to{\rightarrow}

\newcommand{\be}{\begin{equation}}
\newcommand{\ee}{\end{equation}}
\newcommand{\bea}{\begin{eqnarray}}
\newcommand{\eea}{\end{eqnarray}}

\def\ket#1{\left|#1\right\rangle}

\begin{document}

\begin{titlepage}

\bigskip
\bigskip
\bigskip
\bigskip

\begin{center}

{\bf{\Large Quantum Field Theory of Open Spin Networks and New
Spin Foam Models}}

\end{center}
\bigskip
\begin{center}
 A. Mikovi\'c \footnote{Based on the talk given at the X-th Oporto Meeting on Geometry,
 Physics and Topology, Porto, September 20-24, 2001. E-mail address: amikovic@ulusofona.pt}
\end{center}
\begin{center}
 Departamento de Matem\'atica e  Ci\^encias de
Computa\c{c}\~{a}o, Universidade Lusofona, Av. do Campo Grande
376, 1749-024 Lisboa, Portugal
\end{center}

\normalsize

\bigskip
\bigskip

\begin{center}
                        {\bf Abstract}
\end{center}
We describe how a spin-foam state sum model can be reformulated as
a quantum field theory of spin networks, such that the Feynman
diagrams of that field theory are the spin-foam amplitudes. In the
case of open spin networks, we obtain a new type of state-sum
models, which we call the matter spin foam models. In this type of
state-sum models, one labels both the faces and the edges of the
dual two-complex for a manifold triangulation with the simple
objects from a tensor category. In the case of Lie groups, such a
model corresponds to a quantization of a theory whose fields are
the principal bundle connection and the sections of the associated
vector bundles. We briefly discuss the relevance of the matter
spin foam models for quantum gravity and for topological quantum
field theories.
\end{titlepage}
\newpage
\section{Introduction}

The prototype of the state sum models was the Ponzano-Regge (PR)
model of three-dimensional Euclidian quantum general relativity
\cite{pr}. The idea was to construct a theory of quantum gravity
by using the simplical decomposition of the space-time manifold
and group theory. The PR state sum was a sum of amplitudes for all
possible colorings of the edges of a given triangulation with the
irreducible representations (irreps) of $SU(2)$. This sum was
divergent, so that a simple, but topologically non-invariant
cut-off procedure was used to regularize it. Twenty years later,
Turaev and Viro discovered a topologically invariant way to
regularize the PR state sum, by using the quantum $SU(2)$ group
\cite{tv}. Their motivation was not quantum gravity, but they
wanted to find new three-manifold invariants.

Since the Turaeev-Viro (TV) state sum gave new three-manifold
invariants, it was natural to generalize the TV construction to
four-manifolds, which was done by Crane and Yetter \cite{cy,cky}.
Unfortunately, the Crane-Yetter (CY) state sum does not yield  new
manifold invariants. However, by modifying the CY state sum model,
Barrett and Crane have obtained a non-topological model which
describes a theory of four-dimensional quantum gravity
\cite{bce,bcl}. This is an interesting model since there are
indications that it could have the general relativity as a part of
its classical limit \cite{bw,dpf}. Moreover, it has been proven
that a slightly modified BC model \cite{pre,prl}, has finite
amplitudes for regular triangulations \cite{pez,cpr}.

On the other hand, the work on the canonical quantization of the
connection formulation of general relativity, for a review and
references see \cite{lgr}, has revealed the importance of the spin
networks for quantum gravity. It was also realized that the spin
networks are closely related to the spin foams \cite{ba}, while
the spin foams are the basic objects of the quantum gravity
state-sum models \cite{bar}.

The third important development was a realization that the
spin-foam state sum models for Lie groups are the Feynman diagrams
for a partition function of a field theory defined over the group
\cite{boul,oo,dfkr}. Since the usual perturbative quantum field
theories (QFT) have an operator formulation \cite{w}, it was not
difficult to construct the states and the operators which give the
transition amplitudes corresponding to the spin-foam amplitudes
for the manifolds with boundaries \cite{mik}. These boundary
states correspond to the boundary spin networks induced by the
space-time spin foam, and it was shown in \cite{mik2} that
including the open-ended spin nets was equivalent to introducing
matter field operators. In this way one obtains a generalization
of the spin-foam, which we call a matter spin-foam.

In this paper we give a more general description of the results
obtained in \cite{mik,mik2}. We show that these results can be
understood in a simple way from a category theory perspective, and
therefore they can be generalized to the case of tensor categories
associated to quantum groups, or to other tensor categories.

\section{Spin network amplitudes}

Let $\G$ be an oriented graph with a finite set of vertices $V$
and a finite set of edges $E$. We can label each edge $e \in E$
with an irreducible representation (irrep) $\L$ of a Lie algebra
$\bf g$, and each vertex $v \in V$ with an intertwiner operator
$i_v$ corresponding to the irreps of the edges which terminate or
start from $v$. Such a graph is called a spin network, or a spin
net for a short.

In order to be able to make generalizations, for example to the
case of a Hopf algebra $U_q ({\bf g})$, one can give a more
general definition of a spin net. It is a diagram representing a
composition of the morphisms from a tensor category, where the
lines represent simple objects from the category. In the case of
the Lie algebra $\bf g$, the relevant category is the category of
representations of $\bf g$, $Cat({\bf g})$, or an appropriate
tensor sub-category. This interpretation then allows us to
associate a set of complex numbers to a spin net, which we will
call a spin net amplitude, simply by evaluating the components of
this morphism in the bases $\{ e_\a \}$ for the vector spaces $\L$
associated to the edges of $\G$. In this way we obtain \be A^{\a
...}_{\b...}\,(\G,\L,i) = \sum_{...} C_{...}^{...\,(i_1)}\cdots
C^{...\a ... (i_k)}_{........}\cdots C^{.......(i_l)}_{...\b
...}\cdots C_{...}^{...\,(i_V)}\quad,\label{sna}\ee where $\L =
\{\L_1 , ..., \L_E \}$ is the set of irreps of $\bf g$ labeling
the edges of $\G$ and $i =\{i_1 ,....,i_V \}$ is the set of
intertwiners associated to the vertices of $\G$.  The sum in
(\ref{sna}) is over the representation indices of the internal
edges, while $\a ,...$ and $\b,...$ are the representation indices
of the external edges. The numbers $C_{...}^{...(i_v)}$ are the
components of the intertwiner at the vertex $v$. The up indices
represent the domain of a morphism, and the down indices represent
the image of the morphism. One can lower the up indices with the
morphism $C_{\a\b}$, which maps $\L^*$ (dual of $\L$) to $\L$
($e_\a = C_{\a\b} e^\b$), or rise the lower indices with the
morphism $C^{\a\b}$, which maps $\L$ to $\L^*$ ($e^\a = C^{\a\b}
e_\b$). This means that all the internal indices are contracted
(for each down index there is an identical up index).

In order to evaluate (\ref{sna}), it is sufficient to now the
three-vertex amplitude, or the Clebsh-Gordan coefficients
$C_{\a}^{\,\,\,\b\g (i_3)}$. The higher-vertex amplitudes
$C_{...}^{...(i_n)}$, where $n > 3$, are determined by the minimal
three-valent spin net with $n$ external edges.

One can also give a slightly more general definition of the spin
net, by inserting in the edges non-trivial endomorphisms of the
corresponding simple objects. This type of spin nets is obtained
as a basis of square-integrable invariant functions on $G^E$,
where $G$ is a Lie group whose Lie algebra is $\bf g$ \cite{bar}.
In this case the edge endomorphism is given by the representation
matrix $D_\a^\b (g)$ of a group element $g$. Then the spin-net
amplitude (\ref{sna}) is simply the value of the spin-net function
at the point $g_1 =... =g_E = e$ where $e$ is the group identity
element, or a normalized integral of the spin-net function over
$G^E$. This can be generalized to the case of quantum groups
\cite{oeckl}. However, the simpler Lie algebra approach will be
sufficient for our purposes. Still, it will be useful to be aware
of the functions on the group approach in order to understand some
of our constructions.

In the case of the quantum group, we use the category of
representations of the Hopf algebra $U_q ({\bf g})$, so that the
formula for the spin-net amplitude will be again (\ref{sna}), but
now the intertwiner coefficients will be determined by the
Clebsh-Gordan coefficients for the quantum group.

Hence the amplitude for an open spin net is not an invariant, but
it is a tensorial quantity. The case of a closed spin net graph
can be understood as a composition of morphisms between the
trivial one-dimensional irreps. In this case one obtains an
invariant.

The expression (\ref{sna}) will be always finite in the case of
the category of finite-dimensional irreps. In the case of
infinite-dimensional irreps, one may have to do some modifications
to make it finite. The simplest modification is to use an edge
endomorphism $c_\a \d^\a_\b$. For example, the Feynman diagrams of
relativistic particle quantum field theories \cite{w}, are the
spin net amplitudes for the category of unitary representations of
the Poincare group $ISO(3,1)$. These irreps are
infinte-dimensional, and the edge endomorphism is given by the
Feynman propagtor.

In the case of the BC state-sum models, a different type of
modification occurs \cite{bce,bcl}. Since one can think of a spin
net as an invariant function on $G$, one can also consider
invariant functions on the coset space $G/H$, where $H$ is a
subgroup of $G$ \cite{fk}. Consequently one obtains spin nets
where the edges are labelled only by the simple irreps $N$, i.e.
the irreps which have an invariant vector under $H$. Hence the
corresponding spin-net amplitudes can be expressed as multiple
integrals of a propagator \be K_N (x,y) = P_0 D^{(N)}(g_x^{-1}g_y)
P_0 \quad,\ee where $x$ and $y$ are the points in the coset space
$G/H$, which label the vertices of the spin net graph, and $P_0$
is a projector from a simple irrep $N$ onto a trivial irrep of
$H$.

For the BC model, one has $G=SO(4)$ in the Euclidian gravity case,
or $G=SO(3,1)$ in the Lorentzian gravity case, and $H=SO(3)$. In
the Lorentzian case, the unitary irreps of $SL(2,C)$ are used,
which are infinite-dimensional. The Lorentzian spin net amplitudes
require a trivial regulator which is a delta-function of one of
the coset-space coordinates \cite{bcl}.

The coset space construction can be generalized to the case of
open spin nets \cite{mik2}. Let $\L$ be an irrep of $G$ which
contains a finite-dimensional irrep $\t$ of $H$. Then one can
define a matrix function $K_{\L,\t} (x,y)\in End(\t,\t)$ on the
coset space $G/H$ \cite{ch}. Let $\G$ be a spin net graph labelled
with the irreps $\t$ and the corresponding intertwiners $i$. One
can label each vertex of $\G$ with points from $G/H$, and
associate to each internal edge a propagator $K_{\L,\t}(x_j , x_k
)$. Then a spin-net amplitude is given by \be A(\G,\L,\t, i) =
\int \prod_{v\in V} dx_{v} \prod_{v\in V} C_{...}^{...
(i_v)}\prod_{e\in E_{int}} K_{\L_e ,\t_e}
(x_{v_1(e)},x_{v_2(e)})\quad,\label{csa}\ee where $E_{int}$ is the
set of internal edges of $\G$.

\section{Spin foam amplitudes}

The state sum models which had been used for constructing
$d$-dimensional quantum gravity and topological models were based
on a two-complex $J=(V,E,F)$, consisting of a finite number of
vertices $v\in V$, edges $e \in E$ and faces $f\in F$ \cite{bar}.
The two complex $J$ can be based on an arbitrary graph $(E,V)$,
but it is usually taken to be the dual two-skeleton of a
triangulation of the spacetime manifold $M$. One then labels the
faces of $J$ with the irreps of $G$, or with the irreps of the
$q$-deformed $G_q$, while the edges of $J$ will be labelled with
the interwiners for the irreps of the faces that meet at a given
edge. One can regard this object as a generalization of the spin
net, where the one-complex $\G =(V,E)$ is replaced by a set
$(E,F)$ from the two complex $J$, so that  $J$ is called a spin
foam \cite{ba}.

An amplitude for a spin foam can be written as \be A(J) =
\sum_{\L,i} \prod_{f\in F} A_{2} (\L_f) \prod_{e\in E} A_{1}
(\L_{f_1 (e)}\cdots \L_{f_d (e)})\prod_{v\in V}A_{0} (\L_{f_1
(v)}\cdots \L_{f_l (v)}) \quad,\label{sfa}\ee where the sum is
over the labelling set of the irreps and the intertwiners. $A_2$
is the amplitude for a face, and $A_2$ depends on the irrep of
that face. $A_1$ is the amplitude for an edge, and $A_1$ depends
on the irreps of the faces which share that edge. $A_0$ is the
amplitude for a vertex of the two complex $J$, and $A_0$ depends
on the irreps of the faces which intersect at that vertex.

One can choose the amplitudes $A_{0,1,2}$ to be the amplitudes for
the spin nets one can associate to the faces, edges and the
vertices of $J$. A simple way to do this is to replace each edge
in the graph $(V,E)$ by a bundle of $d$ parallel lines, and then
to connect at each vertex of $J$ the $d+1$ bundles of lines into a
one-skeleton graph of a $d$-simplex. In this way, each face of $J$
will correspond to a closed smooth loop of the "bundle" graph.
Hence we can associate to each line in the "bundle" graph an irrep
$\L$. Therefore $A_2$ will be the amplitude for the single loop
spin net, which is $dim\, \L$ for the finite irreps of $G$, or
$dim_q \L$ for the irreps of $G_q$. $A_0$ will be the amplitude
for the $d$-simplex spin net, which will also depend on $d$
intertwiners. For the edge amplitude there is some freedom, i.e.
it could be one, or a given function of $\L_1,...,\L_d$.

In analogy to the spin net case, one can think of a
$d$-dimensional spin foam as a composition of morphisms in a
category where the intertwiner morphisms $i_d$ from $Cat({\bf g})$
belong to the set of  objects. Hence a spin foam will be described
by the $(d+1)$-valent graph, where the edges are labelled by the
morphisms $i_d$, and each vertex will represent the elementary
spin-foam morphism among $d+1$ objects $i_d$. More generally, a
spin foam is a spin net in a 2-category whose objects are the
elements of a tensor category.

The expression (\ref{sfa}) is generically not finite for the
$Cat({\bf g})$ case. When it is finite, then it defines a
topological theory, which happens in $d=2$ compact $G$ case
\cite{bar}. Compact $G$ spin-foam amplitudes are not finite in
$d=3$ and $d=4$ dimensions. In the quantum group case, when $q$ is
a root of unity then the corresponding tensor category has only a
finite number of simple objects. Hence the corresponding spin-foam
amplitude is finite. This gives a topological theory, i.e. the
amplitude is invariant under the Pachner moves. TV and CY
state-sum models are examples of such topological theories.

In analogy to the coset-space spin networks, one can also
construct the coset-space spin foams, where the faces are labelled
by the simple irreps, like in the case of the BC models
\cite{bce,bcl}. BC spin-foam amplitudes can be made finite by
using the edge amplitude $A_1 = A(\Theta_4)$, where $\Theta_4$ is
the theta graph with four edges \cite{pez,cpr}. These models are
relevant for quantum gravity, and they are not topological,
because the simple irreps do not form a tensor category.

\section{QFT of closed spin nets}

In order to evaluate the spin-foam amplitude (\ref{sfa}), we used
the "bundle graph" technique. However, this is the same as the
Feynman diagram formalism from QFT which was developed in the
field theory over the group approach \cite{boul,oo,dfkr,mik}.

Let us describe this in the $d=4$ case. Consider the action \bea S
[\f]&=& \frac12 \,\sum_{\L,i} Q_{\vec\a_1 \vec\a_2}(\times_1,
\times_2)\f^{\vec\a_1}_{\times_1} \f^{\vec\a_2}_{\times_2} +
{1\over 5!}\sum_{\L ,i} \cv_{\vec\a_1 \cdots \vec\a_5}
(\times_1,\cdots ,\times_5)\,\f^{\vec\a_1}_{\times_1}\cdots
\f^{\vec\a_5}_{\times_5}\nonumber\\ &=& S_0 + S_I
\quad,\label{fta}\eea where the complex numbers
$\f^{\vec\a}_{\times} = \f^{\a_1 \cdots \a_4 }_{\L_1 \cdots \L_4
(i)}$ are tensors transforming like $C^{\a_1 ... \a_4 (i)}$. In
the group theory approach, the modes $\f$ are the Fourier modes of
the field $f(g_1 ,...,g_4)$, but in the general case these are
tensors transforming like intertwiners. The vertex functions $Q$
and $\cv$ are given by \bea Q^{-1}(\times_1, \times_2) &=&
\d (i_1 ,i_2)A_1(\L_1,\cdots,\L_4 ) \\
\cv (\times_1,\cdots,\times_5) &=& \d^{10} A_0
(\L_1,\cdots,\L_{10};i_1 ,\,...\,,i_5)\quad,\eea where $\d$ are
the appropriate delta functions of the indices.

The propagator $Q^{-1}$ defines the edge endomorphism $i_1 \to
i_1$ and $\cv$ defines the morphism $i_1 \otimes... \otimes i_5
\to {\bf C}$. The amplitude for a composition of these morphisms
is the same as a Feynman diagram for the perturbative Green's
functions for the path-integral \be \int \prod_i \, d\f_i \,\exp
(-S[\f])\quad.\ee In this way one obtains exactly (\ref{sfa}).

Alternatively, the Fourier modes $\f^{\vec\a}_{\times}$ can be
promoted into creation and annihilation operators \cite{mik}, such
that \be [ \f^{\vec\a}, \f^\dag_{\vec\b}] =
\d^{\vec\a}_{\vec\b}\quad,\ee where $(\f^\dag_{\vec\a})^\dag
=\f^{\vec\a}$. These operators act in the Hilbert space with a
basis
$$ \{ \ket{0} , \f^\dag \ket{0}, \f^\dag \f^\dag \ket{0}, \cdots \}\quad,$$
where the vacuum is defined by $\f^{\vec\a} \ket{0} = 0$. The spin
foam amplitude can be represented as an matrix element of an
evolution operator $(S_I)^n$, where $n$ is the number of the
spin-foam vertices, in analogy with the particle field theory
case. As shown in \cite{mik}, one can construct the ``in'' and the
``out'' states which describe the spatial spin nets corresponding
to the boundary triangulations of the spacetime manifold. Such
spin net states can be represented by \be \ket{\g}=\prod_{v\in
V(\g)} \f^\dag_{\times_v } \ket{0}\quad, \label{snst}\ee where
$\g$ is a four-valent spin net graph dual to the boundary
triangulation, and all the representation indices $\a_v$ are
appropriately contracted.

In the quantum group at root of unity case, one has a well-defined
topological QFT . The formula (\ref{snst}) then gives a map from a
boundary manifold to a Hilbert space. This TQFT should satisfy a
generalization of Atiyah's axioms for manifolds equipped with a
principal $G$-bundle.

\section{QFT of open spin nets}

The construction in the previous section describes only the states
associated to closed spin nets. In the Lie group case it is known
that the closed spin nets are related to the connections on the
principal bundle $(M,G)$. The matter fields can be considered as
sections of the associated vector bundles $(M,\L_s)$, where $\L_s$
is a finite-dimensional representation of $G$. Then it is natural
to consider an open spin net, where a free edge at a vertex $v$
carrying the representation $\L_s$ can be interpreted as a matter
field of "spin" $\L_s$ at $v$. Hence the transition amplitude
between two open $(d-1)$-dimensional spin nets will be described
by a $d$-dimensional spin foam interpolating between them. This
spin foam can be thought of as an morphism in a category where the
objects are the intertwiners $i_d$ and the matter irreps $\L_s$.
Therefore this new morphism, which we will call a matter spin foam
\cite{mik2}, will be described by a graph where the edges are
labelled by $i_d$ or by $\L_s$.

We will only consider the graphs where the sub-graphs formed by
the edges labelled by $i_d$ correspond to triangulations of the
space-time manifold and the edges labelled by the matter irreps
connect only the vertices of that subgraph. This corresponds to
the physical requirement that the matter fields always propagate
in a space-time background. Hence the basic spin-foam vertices
will have $d+1$ edges labeled with the intertwiner morphisms $i_d$
and $n$ edges labelled by the matter irreps of $G$. The
corresponding amplitude will be determined by the spin net with
$n$ external edges labelled by the matter irreps, while the
internal edges labelled by the background irreps will form a
$(d+1)$-simplex spin net.

Let us consider the case of a single matter irrep $\L_s$ and
$d=4$. In the first quantization formalism \cite{mtr,bk}, the
matter can be introduced by replacing the spin net function $\g$
with an open spin net function $\g_s$, which is obtained from $\g$
by putting an external edge carrying a matter irrep $\L_s$ at each
site of $\g$ where a matter quantum is located.

In the QFT formalism \cite{mik2}, one can construct the state
$\ket{\g_s}$ by introducing the matter creation and annihilation
operators $\psi_\s (\times_v)$ and $\psi_\s^\dagger (\times_v)$,
where $\s$ is the representation index of $\L_s$, and the label
$\times_v$ denotes the intertwiner of the spin net site where the
matter quantum is located. Hence the corresponding state would be
given as \be \ket{\g_s}=\prod_{v^\prime \in
V^\prime(\g)}\prod_{\s_{v^\prime}}\psi^\dag_{\s_{v^\prime}}(\times_{v^\prime})
\prod_{v\in V(\g)} \f^\dag_{\times_v}\ket{0} \,,\label{fsn}\ee
where $V^\prime$ is the set of vertices where the matter quanta
are located.

The corresponding matter spin foam amplitude can be expressed as a
matrix element in a QFT  determined by the action $S[\f] + S_s
[\psi,\f]$, where \be S_s = \sum_{\L}
\psi_{\s}(\times)\psi_{\s^\prime}(\times^\prime ) \cv^{\s
\s^\prime} (\times ,\times^\prime , \times_1,\cdots, \times_5 )
\f_{\times_1} \cdots \f_{\times_5} + (h.c.)\quad,\ee and $(h.c.)$
stands for the hermitian conjugate term. Hence \be \cv^{\s
\s^\prime} (\times ,\times^\prime , \times_1,\cdots, \times_5 ) =
\d^{10} \d_{\times,\times_1} \d_{\times^\prime,\times_k} A^{\s
\s^\prime}(\L_1,...,\L_{10})\,,\ee where $A^{\s \s^\prime}$ is a
spin net amplitude for the pentagram with two external edges
attached at the vertices $1$ and $k$, where $1\le k \le 5$. In the
quantum gravity models \cite{mik2}, this graph is zero, and one
needs to consider a modified graph with an extra internal matter
edge between the vertices $1$ and $k$ ($k\ne 1$).

We will also add to $S_s$ a purely quadratic matter term \be
\sum_\L \psi_\s (\times)Q^{\s \s^\prime}(\times ,\times^\prime )
\psi_{\s^\prime} (\times^\prime) + (h.c.)\quad,\ee in order to
have a well-defined perturbative expansion. The propagator
$Q^{-1}$ will be determined by the endomorphism $i_\times \otimes
\L_s \to i_\times \otimes \L_s$. We denote the propagator
amplitude as $G_{\s \s^\prime}$.

Note that $(\psi_\s)^* = \psi^\s =
C^{\s\s^\prime}\psi_{\s^\prime}$, and the vector space duality $*$
is in general different from the complex conjugation (reality)
properties. The reality properties determine relation between the
creation and annihilation operators $\psi$ and $\psi^\dagger$, see
\cite{mik}.

Hence a relevant Feynman diagram $\G_s$, which is generated by the
action $S + S_s$, will be determined by a spacetime skeleton
diagram $\G$ (a 5-valent graph dual to a spacetime triangulation)
plus a line connecting the vertices of $\G$. Therefore $\G_s$ will
be a graph consisting of 5-valent and 7-valent vertices. Out of
these diagrams we will consider only those for which the matter
irreps form a line which connects the centers of a string of the
adjacent 4-simplices, extending from the initial to the final
boundary. In this case, the matter spin foam $J_s$ will be given
by the usual spin foam $J$ and a line $L$ of matter edges with two
external edges. The corresponding amplitude will be given by \be
A_{\s_i}^{\s_f}(J_s ) = \sum_{\L}\prod_{f\in F} A_f (\L)
A_{\s_i}^{\s_f}(L)\prod_{e \in E^\prime}A_e(\L)
 \prod_{v \in V^\prime} A_v(\L) \,,\ee
where \be  A_{\s_i}^{\s_f}(L)= A_{\s_i}^{\s_1}(v_1)\,
G_{\s_1}^{\s_2}(e_{12})\,A_{\s_2}^{\s_3}(v_2)\, \cdots
\,G_{\s_{k-1}}^{\s_k}(e_{k-1,k})\,A_{\s_k}^{\s_f}(v_k)\quad.
\label{fbma}\ee The number of the initial and the final matter
quanta has to be the same for a free theory, and the above formula
can be then easily generalized to the case when there are several
lines $L$, or a loop of matter edges.

One can also take a set of different irreps $\L_{s_1}=S_1$,...,$
\L_{s_k}=S_k$ to represent fields of different spins. The
corresponding QFT action can be written as \bea S_m &=& \sum_{S}
\sum_\L \psi_S(\times)\psi_{S}(\times^\prime )Q_{SS}
(\times,\times^\prime)\nonumber\\ &+& \sum_S \sum_\L \psi_S
(\times) \psi_S (\times^\prime) \cv_{SS} (\times ,\times^\prime ,
\times_1,\cdots,
\times_5 ) \f_{\times_1} \cdots \f_{\times_5}\nonumber\\
&+&\sum_S \sum_\L \psi_{S_1}(\times) \cdots
\psi_{S_k}(\times^{(k)}) \cv_{S_1 ... S_k}
(\times,...,\times^{(k)}; \times_1 ,...,\times_5) \f_{\times_1}
\cdots \f_{\times_5} \nonumber\\ &+& (h.c.)\quad,\label{tia}\eea
where the vertex $\cv$ will be determined by the spin net
amplitudes for the open graphs based on the pentagram. A new
feature is that the set of possible graphs for the vertex
$\cv_{S_1 ... S_k}$ will contain graphs with more than five
vertices, i.e. there will be vertices formed by the matter irreps
describing the interactions among the matter fields \cite{mik2}.

The corresponding Feynman diagrams, or the spin-foam amplitudes,
will be generically divergent in the Lie group case. By going to
tensor categories with finitely many simple objects, like those
for quantum groups at root of unity, one will obtain finite
numbers. Since such amplitudes are topologically invariant when
there are no matter irreps, then it will be interesting to explore
the topological invariance of the corresponding amplitudes with
the matter irreps, since one then expects to obtain topological
models.

Another way of obtaining finite expressions is by using the coset
space spin networks. The corresponding models will be relevant for
quantum gravity with matter fields \cite{mik2}, and they will not
be topological. This requires a specification of the irreps $s$ of
$H$ which are contained in the irreps $S$ of $G$. In that case the
matter fields will have the indices from the $H$ irreps, and the
action can be written as \bea S_m &=& \sum_{S} \sum_N
\psi_s(\times)\psi_{s}(\times^\prime )Q^S_{ss}
(\times,\times^\prime)\nonumber\\ &+& \sum_S \sum_N \psi_s
(\times)\psi_s (\times^\prime )\cv^S_{ss} (\times ,\times^\prime ,
\times_1,\cdots,
\times_5 ) \f_{\times_1} \cdots \f_{\times_5}\nonumber\\
&+&\sum_S \sum_N \psi_{s_1}(\times) \cdots
\psi_{s_k}(\times^{(k)}) \cv^{S_1 ... S_k}_{s_1 ... s_k}
(\times,...,\times^{(k)}; \times_1
,...,\times_5) \f_{\times_1} \cdots \f_{\times_5} \nonumber\\
&+& (h.c)\quad,\label{ntia}\eea where $N$ are the simple
background irreps and the vertices $Q$ and $\cv$ will be
determined by the spin net amplitudes for open graphs based on the
$\Th_4$, $\Th_5$ and the pentagram graphs. However, these
amplitudes will be now given as multiple integrals of the
propagators $K_N (x,y)$ and $K_{S,s} (x,y)$, as in the equation
(\ref{csa}).

In the case of the Lorentzian BC model, $G=SO(3,1)$, $H=SO(3)$,
and the background irreps are the simple irreps $N=(0,\r)$, $\r\ge
0$, which are unitary and infinite dimensional. The matter can be
introduced by using the finite-dimensional irreps of $SO(3,1)$
\cite{mik2}, which can be labelled as $(j,k)$, $j,k \in \frac12
{\bf Z}_+$, and these irreps  are non-unitary. To each $S=(j,k)$
irrep, we will associate a $SU(2)$ irrep $s$ contained in $S$.
Hence one will have spin foams where the faces will be labelled
with the $N$ irreps, while the edges will be labelled by the $S$
and $s$ irreps.

\section{Concluding remarks}

Note that by generalizing the notion of the spin network as a
composition of morphisms among simple objects from a tensor
category, one is able to generalize the notion of a spin foam as a
spin net in a 2-category. In this way one obtains matter spin
foams, where the edges of the graph are labelled by the simple
objects or by the intertwiners. If we think of this graph as a
dual one-skeleton of a triangulation, then in the Lie group case,
we label the edges with the irreps of $G$, or with the
intertwiners, while the faces (loops of the graph) are labelled by
the irreps corresponding to the edge intertwiners. It would be
interesting to explore the topological properties of these
amplitudes, especially in the quantum group case, where the
amplitudes are finite. This may give topological invariants for
manifolds equipped with principal and associated vector bundles.

In the case of the BC model with matter, we expect that the
amplitudes are finite, because the amplitudes without matter are
finite, and adding matter introduces the matter propagators
$K_{S,s}$, which are similar functions of the coset space points
as the $K_N$ propagators \cite{mik2}. In addition, one can
consider the amplitude (\ref{fbma}) as an amplitude for matter
propagation in a fixed set of the background irreps, which could
be interpreted as an amplitude of a quantum field theory in a
curved background.

\end{document}